\documentclass[11pt,a4paper]{article}
\usepackage[T1]{fontenc}
\usepackage[utf8]{inputenc}
\usepackage{authblk}

\usepackage{geometry}
\usepackage{amsfonts,amsmath,amssymb,amsthm,nicefrac}
\usepackage{ulem}
\usepackage{cancel}
\usepackage{yhmath}
\usepackage{arcs}
\usepackage{graphicx}
\usepackage{url}
\usepackage{subcaption}
\usepackage{url}
\usepackage{caption}
\usepackage{subcaption}

\title{\bf On the nonclassicality distance indicator of qudits}
\author[1,2,3]{Arsen Khvedelidze}
\author[3]{Astghik Torosyan}

\affil[1]{A. Razmadze Mathematical Institute, Iv. Javakhishvili Tbilisi State University, Tbilisi, Georgia}
\affil[2]{Institute of Quantum Physics and Engineering Technologies, Georgian Technical University, Tbilisi, Georgia}
\affil[3]{Laboratory of Information Technologies, Joint Institute for Nuclear Research, Dubna, Russia}

\date{ }

\begin{document}

\maketitle

\begin{abstract}
We consider the nonclassicality distance indicator of a state in finite-dimensional quantum systems which is evaluating  a state nonclassicality by its  remoteness from the set of ``classical states''.
The latter are identified  with those states whose Wigner function is non-negative.
The corresponding  Wigner function's positivity polytope in the simplex of qudit eigenvalues is introduced and the representation for the nonclassicality distance indicator as a piecewise function is derived. The results are exemplified by the qutrit case. 
\end{abstract} 

\tableofcontents

\newpage 

\label{sec:intro}
\section*{Introduction}

It is commonly accepted that the negativity 
of quasiprobability distributions of quantum states is an essential attribute of ``quantumness''.
Following this understanding, one can identify among quantum states $\varrho \in \mathfrak{P}_N$  a subset of the ``classical states'' $\mathfrak{P}_{\mathrm{Cl}}$, whose quasiprobability distributions are non-negative. The complement of $\mathfrak{P}_{\mathrm{Cl}}$ consists of states - carriers of a certain ``quantumness''. 
To quantify the ``amount of quantumness'' 
in a state $\varrho $ we ascribe the meaning of nonclassicality measure to its distance from a subset of classical states    $\mathfrak{P}_{\mathrm{Cl}} \subseteq \mathfrak{P}_N$: 
\begin{equation}
\label{eq:NCDindic}
\mathrm{d}_{\mathfrak{P}_{\mathrm{Cl}}}(\varrho) =\inf_{x \in \mathfrak{P}_{\mathrm{Cl}}}\,\mathrm{D}(\varrho, x )\,.  
\end{equation}
The idea to use the remoteness of state from the assigned classical subset is borrowed  from quantum optics, where M.Hillery  \cite{Hillery1987}
introduced the nonclassicality measure using the  distance D corresponding to the trace-norm of radiation density matrices. Below a few generic properties of (\ref{eq:NCDindic}) for a finite-dimensional quantum system endowed with  the Frobenius (Hilbert-Schmidt) norm will be formulated. Being restricted by the volume of the publication, we omit all the proofs and concentrate only on the results of our calculations of $\mathrm{d}_{\mathfrak{P}_{\mathrm{Cl}}}(\varrho)$ for a generic 3-level system, the qutrit.

\section{Wigner quasiprobability distribution of qudit}

The Wigner function (WF)  of an $N$\--level quantum system \--- qudit \---  is a pairing of a density matrix $\varrho$ from the state space $\mathfrak{P}_N$ and an element from  
the dual space $\mathfrak{P}^\ast$,  the Stratonovich-Weyl (SW) kernel  $\Delta(\boldsymbol{z})$ 
defined as the matrix valued function over the phase space $\Omega_N$:
\begin{equation}\label{eq:WignerFunction}
W_\varrho(\boldsymbol{z})= \mbox{tr}\left[\varrho\, \Delta(\boldsymbol{z})\right]\,, \qquad  
\boldsymbol{z} 
\in \Omega_N\,.
\end{equation} 
The kernel  $ \Delta(\boldsymbol{z}) $  is a Hermitian solution to the ``master equations''
~\cite{AKh2017-WF}:
\begin{equation}
\label{eq:mastereq} 
{\mbox{tr}\left[\Delta(\boldsymbol{z})\right]=1} \,,  \qquad  
{\mbox{tr}[\Delta(\boldsymbol{z})^2] = N} \,.
\end{equation} 
The group of isotropy $\mathrm{Iso_\Delta}$ of $ \Delta $ with respect to  $U(N)$ action  dictates coset structure of the phase space,
$\Omega_N=\mathrm{U}(N)/\mathrm{Iso_\Delta}\,.$ Non-uniqueness of the solutions to (\ref{eq:mastereq})   
leads to the existence of a  family of $N-2$ unitary non-equivalent Wigner functions, enumerated 
by the moduli parameters $\boldsymbol{\nu} = \left(\nu_1,\ldots,\nu_{N-2}\right)$\,: 
\begin{equation}
\label{eq:WFN}
W^{(\boldsymbol{\nu})}_{\varrho} (\boldsymbol{z})=\frac{1}{N} \left[1 + \frac{N^2-1}{\sqrt{N+1}} \,(\boldsymbol{n}^{(\boldsymbol{\nu})}\,, \boldsymbol{\alpha})\right]\,. 
\end{equation}
In (\ref{eq:WFN})
the Bloch vector of a state $\varrho$ is denoted as 
$\boldsymbol{\alpha},$
a unit vector $\boldsymbol{n}^{(\boldsymbol{\nu})}$ is a function of the coset coordinates $\boldsymbol{z}\in \Omega_N\,$  and moduli parameters $\boldsymbol{\nu}$
(see details in \cite{AKh2017-WF}).
The family of WFs satisfies  all properties of  the statistical probability distributions, except positivity. But we can  pick out the states 
with proper probability distributions, 
\begin{equation}
\label{eq:P+}
\mathfrak{P}_{\mathrm{Cl}} = \{\, \varrho \in \mathfrak{P}_N\,\ | \  W_\varrho(z) \geq 0\,, \quad  \forall z\in \Omega_N \, \}\,,
\end{equation}
and call them ``classical states''.
Bearing in mind the above definitions, one can prove the following assertions:  
\begin{enumerate}
\item[I.] The hyperplane in the simplex of  eigenvalues  of $\varrho \in \mathfrak{P}_N$,
\begin{equation}
\label{eq:hyperplane}
 r_1\pi_{N} +r_2\pi_{N-1}+ \ldots +r_{N-1}\pi_2+r_{N}\pi_1=0\,,
\end{equation}
separates the sets of states whose WFs have opposite signs. 
\item[II.] In  $\mathfrak{P}_N$ there is the
Hilbert-Schmidt ball of ``absolute positivity'' with the center at   $\varrho_0:={1}/{N}\,\mathbb{I}_N\,,$ of radius $r_\ast = \sqrt{N+1}/(N^2-1)$\,,
such that all states inside it are classical for all moduli parameters $\boldsymbol{\nu}$\,;
\item[III.] 
The hyperplane 
(\ref{eq:hyperplane}) is tangent to the ball of ``absolute positivity'' at states with 
\begin{equation*}
\mbox{spec} (\varrho_\ast) = \frac{1}{N^2-1} \left(N-\pi_N\,, N-\pi_{N-1}\,, \dots\,, N-\pi_1\right)\,. 
\end{equation*}
\end{enumerate}
Note that in the statements (I)-(III) it is assumed that the spectra of the density matrix $\mbox{spec}(\varrho)=\{r_1, r_2, \dots, r_N\} \,$ and the SW  kernel $\mbox{spec}(\Delta)=\{\pi_1, \pi_2, \dots, \pi_N\},$ are both ordered in a decreasing  way.

\section{Nonclassicality distance of qutrit} 

Exploiting properties (I)-(III), we were able to derive an exact expression for the nonclassicality distance of a generic state of a 3-level  system for an arbitrary representation of WFs.  Briefly the results can be formulated as follows.

\begin{figure}[h!]
\begin{minipage}[h]{0.32\linewidth}
\center{\includegraphics[width=0.9\linewidth]{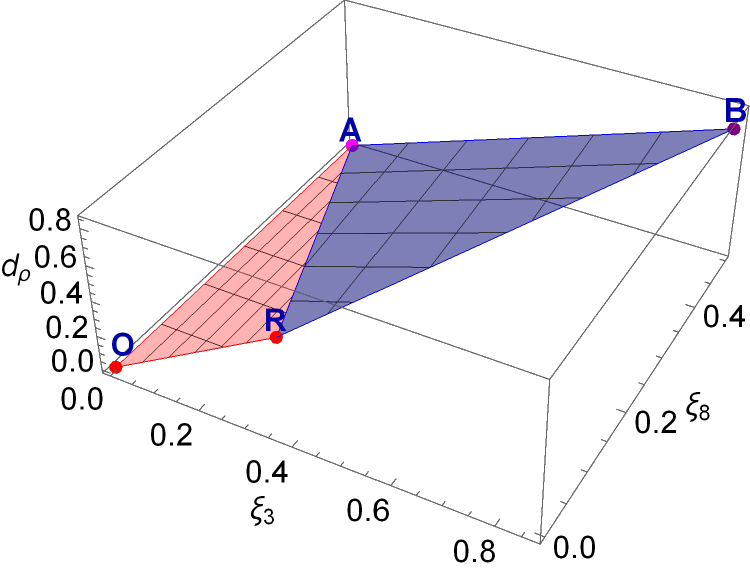}}
\end{minipage}
\hfill
\begin{minipage}[h]{0.32\linewidth}
\center{\includegraphics[width=0.9\linewidth]{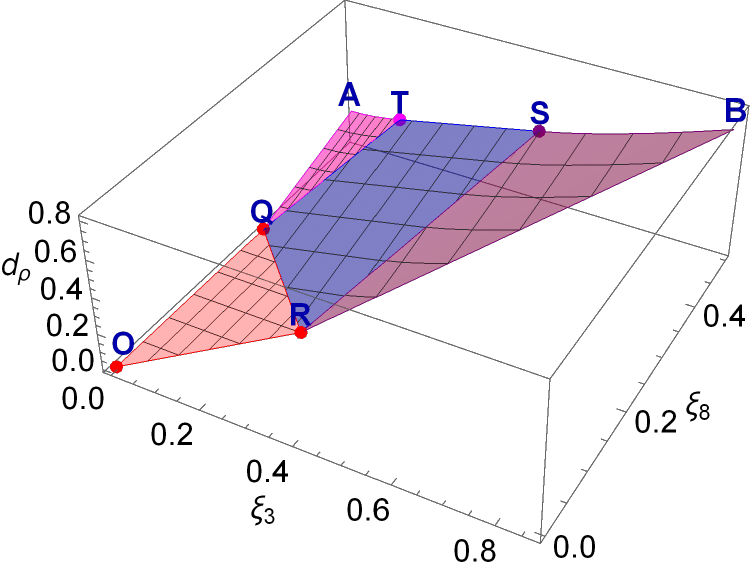}}
\end{minipage}
\hfill
\begin{minipage}[h]{0.32\linewidth}
\center{\includegraphics[width=0.9\linewidth]{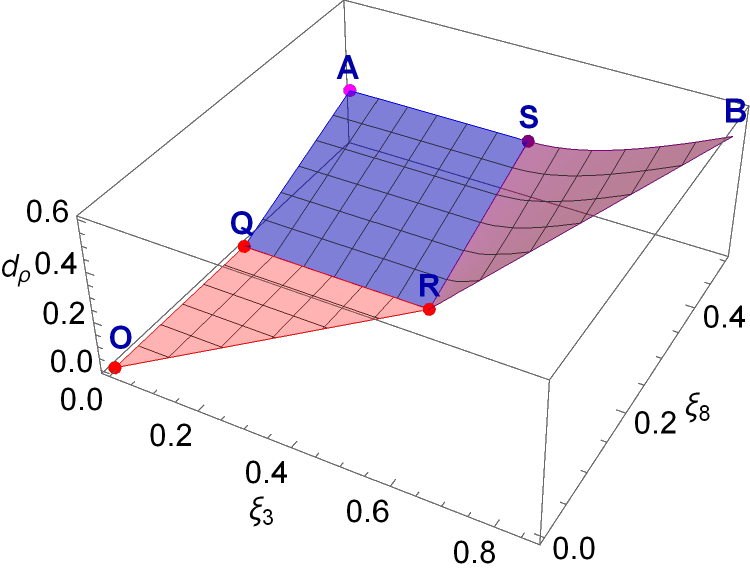}}
\end{minipage}
\begin{minipage}[h]{0.96\linewidth}
\begin{tabular}{p{0.32\linewidth}p{0.32\linewidth}p{0.32\linewidth}}
\centering 
\footnotesize $\zeta=0$ & \centering 
\footnotesize $\zeta=\pi/6$ & \centering 
\footnotesize $\zeta=\pi/3$ \\
\end{tabular}
\end{minipage}
\caption{Qutrit orbit space, the triangle  in $(\xi_3,\xi_8)$ plane, divided into WF's positivity triangle  and  supports of 
$\mathrm{d}_{\mathfrak{P}_{\mathrm{Cl}}}(\varrho)$ 
for different values of the moduli parameter $\zeta\,.$}
\label{fig:qutrit-nc-distance-simplex}
\end{figure}

\begin{figure}[h!]
\begin{minipage}[h]{0.32\linewidth}
\center{\includegraphics[width=0.9\linewidth]{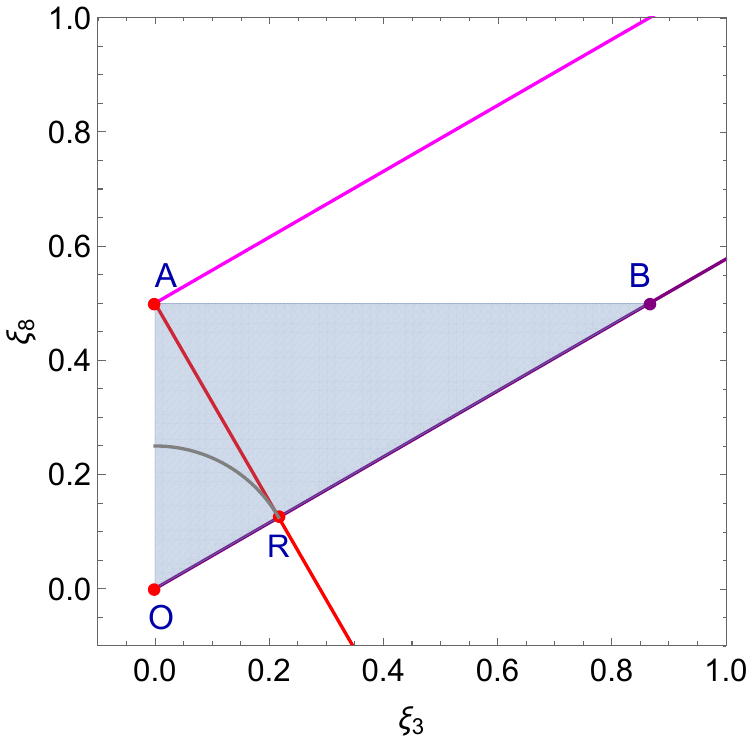}}
\end{minipage}
\hfill
\begin{minipage}[h]{0.32\linewidth}
\center{\includegraphics[width=0.9\linewidth]{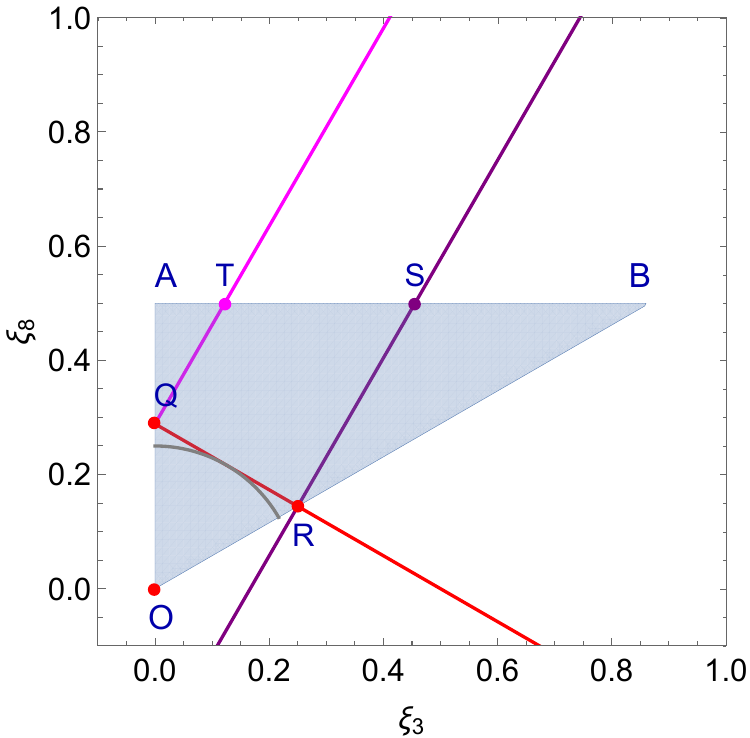}}
\end{minipage}
\hfill
\begin{minipage}[h]{0.32\linewidth}
\center{\includegraphics[width=0.9\linewidth]{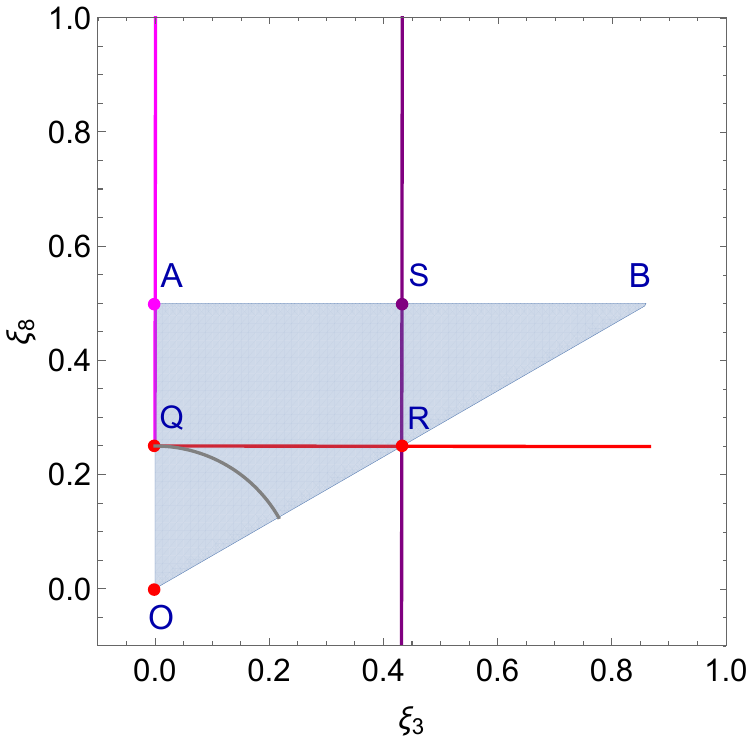}}
\end{minipage}
\begin{minipage}[h]{0.96\linewidth}
\begin{tabular}{p{0.32\linewidth}p{0.32\linewidth}p{0.32\linewidth}}
\centering 
\footnotesize $\zeta=0$ & \centering 
\footnotesize $\zeta=\pi/6$ & \centering 
\footnotesize $\zeta=\pi/3$ \\
\end{tabular}
\end{minipage}
\caption{Qutrit nonclassicality distance  for different values of moduli parameter.} 
\label{fig:qutrit-nc-distance}
\end{figure}

According to the ``master equations'' (\ref{eq:mastereq}),  the spectrum of a qutrit SW kernel can be parametrized by $\zeta \in [0, \pi/3]$ \--- the angle specifying  the representative  of admissible qutrit WFs, 
\begin{equation}
\mbox{spec}(\Delta)= \frac{1}{3}\{ 1+2\sqrt{3}\sin{\zeta}+2\cos{\zeta}\,, 1-2\sqrt{3}\sin{\zeta}+2
\cos{\zeta}\,, 1-4\cos{\zeta} \}\,.
\end{equation}
Calculating (\ref{eq:NCDindic}), we use a generic qutrit density matrix in the following form:
\begin{equation} 
\label{eq:QutritDM}
\varrho = 
\frac{1}{3}\mathbb{I}_{3} + \frac{1}{\sqrt{3}}\,\sum_{i=3,8} \xi_i\,
U\,\lambda_i\, U^{\dagger}\,,
\end{equation}
with the unitary factor 
$U\in SU(3)\,$ 
and real coefficients  $\xi_3, \xi_8 \in \triangle{OAB}$ (see  Fig.\ref{fig:qutrit-nc-distance-simplex}) which are 
$SU(3)$\--invariants characterizing the state  $\varrho$. 
\footnote{
In (\ref{eq:QutritDM}) the Gell-Mann basis $\boldsymbol{\lambda}=\{\lambda_1,\ldots, \lambda_{8}\}$ of  $\mathfrak{su}(3)$ algebra,  $\mbox{tr}(\lambda_i\lambda_j)=2\delta_{ij}$,  is used.}
As a result, we arrive at the following piecewise function of the nonclassicality distance  indicator based on the Hilbert-Schmidt distance:
\begin{equation*}
\mathrm{d}_{\mathfrak{P}_{\mathrm{Cl}}}(\varrho)= \begin{cases}
    0\,, \hfill  \mbox{if} \quad \xi_3,\xi_8 \in \triangle{OQR}\,, \\ 
\sqrt{\xi_3^2+\left(\xi_8-\frac{1}{4}\sec (\zeta-\frac{\pi}{3})\right){}^2}\,, \hfill \mbox{if} \quad \xi_3,\xi_8 \in \triangle{AQT}\,,
    \\ 
\xi_3\cos{\left(\zeta+\frac{\pi}{6}\right)} + \xi_8\sin{\left(\zeta+\frac{\pi}{6}\right)}-\frac{1}{4}, \hfill \mbox{if} \ \xi_3,\xi_8 \in \square{QRST}\,, \\ 
    \sqrt{\left(\xi_3-\frac{\sqrt{3}}{8}\sec(\zeta )\right)^2+\left(\xi_8-\frac{\sec(\zeta)}{8}\right)^2}\,, \hfill \mbox{if} \quad \xi_3,\xi_8 \in \triangle{BRS}\,.
\end{cases}
\end{equation*} 
The triangle  of WF positivity,  ($\triangle{OAR}$ or $ \triangle{OQR}$), disk of absolute positivity and support of $\mathrm{d}_{\mathfrak{P}_{\mathrm{Cl}}}$ for $\zeta=(0,\frac{\pi}{6}, \frac{\pi}{3}) $ are depicted in 
Figs.\ref{fig:qutrit-nc-distance-simplex}, \ref{fig:qutrit-nc-distance} respectively. 

\section{Conclusion }

Searching for a minimum  in (\ref{eq:NCDindic}), we introduce a linear functional $\mathfrak{P}_N \to \mathbb{R}$: 
\begin{equation}
    w[\varrho] := \inf_{g \in U(N)} \, W_{g \varrho g^\dagger}\, (z)\,,
\end{equation}
whose zero level, 
$ w [\varrho] = 0,  
$ describes  the supporting hyperplane (\ref{eq:hyperplane})
of the  convex set of classical states
$\mathfrak{P}_{\mathrm{Cl}}.$ 
The hyperplane cuts out the convex WF positivity polytope from the simplex of qudit eigenvalues.
A knowledge of the WF positivity  polytope allows one to extract information on the quantumness of states  not only from the nonclassicality distance indicator (\ref{eq:NCDindic}) but from other  nonclassicality measures as well. Particularly, the results of its usage in calculations of such  
measures as the global indicator of nonclassicality 
\cite{AKhTGI2020} and the Kenfack-$\dot{\text{Z}}$yczkowski indicator
\cite{AKhT-KZ2021} will be 
given in the forthcoming publications.


\end{document}